# Consciousness is not a physically provable property

Catherine M Reason[1]

*We present a logical proof that computing machines, and by extension physical systems, can never be certain if they possess conscious awareness.  This implies that human consciousness is associated with a violation of energy conservation.  We examine the significance that a particular interpretation of quantum mechanics, known as single mind Q (Barrett 1999), might have for the detection of such a violation. Finally we apply single mind Q to the problem of free will as it arises in some celebrated experiments by the neurophysiologist Benjamin Libet.*

In 1995 Gilbert Caplain published a paper entitled "Is consciousness a computational property?", in which he outlined an argument to the effect that no computing machine could ever be conscious.  In his paper, Caplain pointed out that his argument was presented only in outline, and that some of the ideas presented required further work (Caplain 1995, 2000).  In this author's opinion Caplain's argument is not, in fact, an argument that consciousness is not a computational property but rather something more subtle; it is an argument that no computing machine can ever, using purely computational processes, be certain if it is conscious.

To establish his argument, Caplain demonstrates an inconsistency between two principles; the principle of *reflexivity* and the principle of *cognitive separation. Reflexivity* is Caplain's term for the capacity of conscious beings to know with certainty that they are conscious; *cognitive separation* can be expressed as the separation between some symbolic state in a computing machine, and the state of affairs which that state represents.  Caplain argues that, if all computing machines are bound by the principle of cognitive separation, then the inconsistency between these two principles implies that no computing machine can ever be truly conscious, and hence conscious human beings cannot be computing machines.  This argument effectively applies Descartes' notion of the malicious genius to the internal states of a computing machine.

It seems to this author that Caplain's use of the term *reflexivity* does not conform to the usual philosophical usage, and so I shall use the term *self-certainty* instead.  To

[1]To whom correspondence should be addressed at CMRneuro@Gmail.com

avoid any ambiguity we shall define this term here:

Definition: *Self-certainty* is the capacity of at least some conscious beings to verify with certainty that they are conscious.

The detailed proof of Caplain's result that we are presenting here is substantially different from Caplain's in form, and attempts to minimize any dependence on philosophically ambiguous terms such as "knowledge" and "belief". However it relies on the same properties of consciousness and of machines. For the purposes of this argument, a computational process is operationally defined as any process which can be represented in the following form:

$$\text{Result} = P(\text{input})$$

where P is some computation. The exact form of P itself is irrelevant to this argument, so according to this definition a computational process is any computation which associates an input to an output. A computation here means simply any process which occurs in a computing machine. If the reader is concerned that this leaves the term "computing machine" undefined, then this may be taken to mean "some Turing machine", although this is not in fact a necessary stipulation.

In order to show that no computing machine can verify with certainty that it is conscious, one must first assume a computing machine M, all of whose computations are assumed to take the form above. At this point we must also define the following Principle F (the functionalist principle):

"Every human mental process supervenes on some computational process."

This principle asserts, in effect, that human beings are computing machines of the same form as M. M is now presented with the task with the task of proving that it is conscious. At this point two conditions must be noted:

1  M is given the task of proving that it is *certainly* conscious. Proofs that M may be conscious which depend on additional assumptions, or which fall within particular limits of confidence short of full certainty, fall outside the scope of this argument and are not relevant to it.

2  "Conscious" in this context, does not necessarily mean "awake" or "self-conscious". It means only that some form of conscious experience is present, even if this is some altered state of consciousness such as a lucid dream. (It may seem odd to attribute such states to machines, but as it is impossible to assert, a priori, what forms consciousness may take in computing machines, this possibility must be allowed for.)

At this point the reader should be careful to attend to the following operational definitions. Firstly we operationally define *certainty* as follows: M is *certain* of some proposition k if M is able to determine that k is certainly true. Other definitions of certainty -- for example, subjective "feelings" of being certain -- are not relevant to this argument. Secondly we operationally define *provable*, in statements of the form "proposition k is provable by M" as meaning: M is able to determine that k is certainly true. The reader should be careful not to confuse this operational definition with more familiar notions, for example those concerning the proof of theorems in formal systems.

M's task can now be represented as a function or mapping from a domain E to a range X. E is a binary variable which represents the presence or absence of conscious experience and takes the following values:

$E = 1$  if conscious experience is present when the mapping is performed;

$E = 0$  if no conscious experience is present.

X is a binary variable which takes the following values:

$X = YES$  if $E = 1$

$X = NO$  if $E = 0$ or if the state of E cannot be ascertained.

The mapping therefore associates a state E, which represents the presence or absence of consciousness, with a state X which represents the answer to the question "Am I conscious?" This mapping is performed by a computation P which can be represented as follows:

$X = P(E)$

where X and E can now be thought of as states (or sets of states) in M. It is necessary also to make the following assumptions:

1  M can reason deductively (in particular, M must have deductive reasoning powers equivalent to those of a human being). It is not necessary to specify exactly what these powers are; merely that there is an equivalence between humans and M.

2  M is "honest" -- that is, there are no systematic biases which prevent M from reasoning deductively in the domain in question. This is actually quite a difficult requirement to make precise. The best approach is to assert that there are no systematic biases which would make it impossible, even in principle, for M to follow classical rules of inference such as modus ponens.

We now define the following deductive argument which I shall call A:

The reliability and accuracy of the computational process $X = P(E)$ depend critically on the reliability and accuracy of P (which is to say, how well P performs the mapping from E on to X). Consider some malformed computation BadP such that

$X = BadP(E = 0) = YES$

In such a case, M will conclude that it is conscious, but M's conclusion will be neither accurate nor reliable. Therefore the accuracy of P needs to be checked, and by Principle F, this must be done by some computation P', such that

$$X' = P'(P)$$

where X' is YES if P is found to be accurate and NO otherwise. But what of the reliability and accuracy of P'? Clearly this would necessitate some further computation P'' to establish the accuracy of P' -- and so on leading to an infinite regress. It follows that the reliability and accuracy of P can never be ascertained with certainty, and hence the value of E cannot be ascertained with certainty either. (One can paraphrase this by saying that, in any system which relies entirely on computations, the reliability and accuracy of any given computation can only be determined by applying another computation to it, and this process is obviously non-terminating.) It should be noted here that this argument applies even if P = P' (that is, if P and P' are the same process) since it does not follow that X = X'. (As the input is different, the output can be different even if the function is the same.)

It follows from this that X cannot be guaranteed to be reliable indicator of the value of E, and nor can the value of any subsequent state, such as X', render X ultimately reliable as an indicator of the value of E. In plain language this means that X, which represents M's answer to the question "Am I conscious?", can never be relied upon to be a certainly correct answer to that question, so long as the value of X is determined by some computation. It is not possible, by means of any computation, to establish with certainty the value of E, and since M is a computing machine, M can never establish with certainty that it is conscious. This concludes the definition of Argument A.

It follows from Assumptions 1 and 2 that M can deduce A, and thereby deduce that it can never be certain if it is conscious. This rules out the possibility that M could be conscious, and arrive at the correct conclusion that it is conscious via faulty reasoning. Given our assumptions, it is simply impossible for M to be certain that it is conscious. It is important to note the two stages of this process. Argument A simply implies the potential *unreliability* of M (M may be accurate but it is impossible to establish this with certainty by means of any computation). Assumptions 1 and 2 allow M to deduce A and thus deduce the *uncertainty* of M (M can show that it can never be certain of the accuracy of any of its computations). (Incidentally it is not necessary for M to assume that it is a computing machine; it is sufficient for M to be unable to establish with certainty that it is not a computing machine.) This argument has a recursive character which may seem a little baffling at first sight, since the reader's brain is itself part of the argument! That is, we rely on the reader to appreciate the soundness of the deductive argument A. Once this is given, then Assumption 1 guarantees that M will also appreciate the soundness of A.

It is now apparent that M cannot possess self-certainty. But conscious human beings *do* possess self-certainty; it is possible for a conscious human being to know, with absolute certainty, that they are conscious (in the sense defined above in Condition 2). This implies that Principle F (which asserts that human beings are computing machines of the same form as M) must be wrong. It is in this sense that we can say that consciousness is not a computational property -- or that if it is, it is attended by some other property or properties which are not themselves computational in nature. At this point it should be remembered that this proof applies only if M possesses deductive powers similar to those of a human being (Assumption 1). Conceivably if M did not possess such powers, then M could not deduce the argument A, and the proof of M's uncertainty would not apply; however in such a case, human beings could not be machines of the same form as M.

**Expansion of the computational argument to physical processes**

In the previous section P was considered to be a computation mapping E on to X. However there is no reason to confine the definition of P in this way. P can instead be regarded as any *physical process* which performs the same mapping, and M can be regarded as a *physical system* rather than specifically a computing machine. To eliminate any confusion between mappings, computations, and physical processes, the relation between P and X can be rewritten to avoid any explicit mention of E:

$$X = O(P)$$

X is a binary variable as before, but P is now a physical process whose output O determines the value of X, where X is some state (or set of states) in a physical system. This formulation is intended to make it clear that physical processes which perform functions or mappings may not in any sense "look like" computations; in other words, they may not take the form of operations on data inputs. Once again, the reader may worry that the term "physical process" is effectively undefined. A physical process can therefore be operationally defined as any objective entity in the real world which has the potential to evolve in time. This includes for example collections of molecules, or computers running programs, but excludes abstract entities such as mathematical functions, or programs without implementations to run them. The output O of a physical process can be regarded as just the effect which that process has on the value of X. A *physical system* can be regarded as some set of physical processes.

It is now also necessary to change the Principle F to the following Principle F' (the physicalist principle):

"All human mental processes supervene on some physical process."

Argument A then proceeds much as before, except that the word "computation" in A is replaced by the word "process". Again one notes the possibility of physical processes BadP such that:

$$X = O(BadP) = YES$$

even when E = 0. This necessitates some physical process P' to ascertain the accuracy and reliability of P, and as before, this leads to an infinite regress.

This is all that is needed to show that, either consciousness is not a physical property, or it is attended by a property or properties which cannot themselves be physical. As before, Assumptions 1 and 2 imply that M can deduce the Argument A, and thereby establish that it can never be certain of being conscious.  The upshot is that any physical system capable of reasoning honestly and which has deductive reasoning powers equivalent to those of a human being, would have to conclude that the question "Am I certainly conscious?" is effectively undecidable.  Consciousness, therefore, is not a physically provable property.

How can this be?   It is an inevitable consequence of the separation between the state of X and the process P by which the state of X is determined.  This is analogous to Caplain's  principle of cognitive separation.  But it can readily be seen that it applies to *any* process P such that X is the output of P.  In fact, even the qualifier "physical" is redundant; this argument applies to any sort of process *whatsoever* if the state of X is determined by the output of that process, rather than directly by E with no intervening process of any sort.[2]

The reader may feel that this limitation on the capabilities of physical systems is too trivial to be worth mentioning.  It simply means that humans beings derive their certainty of being conscious not by any sort of mediating process, but by what in philosophy is called "acquaintance".  However it has a serious consequence which has received virtually no attention within the academic literature.  Principle F' implies that if M cannot be certain that it is conscious, then human beings cannot be certain that they are conscious either.  Principle F' is therefore inconsistent with the property of self-certainty.[3]   So -- either Principle F' is wrong, or one of the other assumptions does not apply to human beings.

It can be noted immediately that Assumption 1 cannot be discarded since by definition it must apply to human beings.  Assumption 2 could be discarded but would leave one with the somewhat paradoxical situation that humans could be certain of being conscious only because their brains were incapable of honest reasoning (and hence were unreliable).  Nonetheless, as we shall see later, there may be situations in which Assumption 2 could at least be modified, though to discard it entirely would be asking rather a lot of coincidence; it would in effect require a faulty system to produce, and produce reliably, the correct result via a series of fortuitous accidents. There could also be no way for humans to establish with certainty that the flaw in their reasoning was precisely that flaw required, for them to reach the correct answer to the question "Am I conscious?"  This seems to leave one with no choice but to throw out Principle F'.  Human mental processes, in other words, do not all supervene on physical processes.[4]

---

[2]In fact it is not enough for E directly to determine X; E must also directly determine *that it is the case* that E directly determines X, and do so in a way that a conscious subject can be certain is reliable -- that is, not by means of any physical process which would be susceptible to Argument A.

[3]Another way of looking at this is to say that knowledge or understanding by "acquaintance" is impossible in any physical system; or that if it is possible, it cannot influence the evolution of that system.

[4]One might think that allowing X to be identical with E might solve this problem -- that is, by allowing X to be a state which is identical with consciousness itself.  It is obviously possible to arrange things so that if E and X are identical, then it must be the case that E = 1 if  X = YES.  But to make use of this (and thus to be certain that X can be relied on)  M must have some way of being certain that it is the case that E and X are identical.  Since M is a physical system, any means of obtaining such a proof

It is important to note that this conclusion applies not only to consciousness itself, but to some of the contents of consciousness as well. It also follows from Argument A that if human beings were exclusively physical systems, they could not be certain of the truth of the statement "I am reading this article"; indeed they could not even be certain of the truth of the statement "I believe I am reading this article". One could even formulate Argument A in such a way that physical systems could not be certain of their own existence.

There is also an important difference between this conclusion concerning consciousness in physical systems, and the original, more restricted conclusion regarding computing machines. This is because even if human beings can be certain that they have conscious experience, it is still the case that physical systems -- such as brains -- cannot. This implies that when human beings ask themselves if they are conscious, either the evolution of their mental processes will diverge from the physical evolution of their brain-states in some drastic and irreversible manner; or their mental processes will force their brains to evolve in a manner which is inconsistent with their own physically determined behavior. Such a violation of physically determined behavior should entail -- at the very least -- a violation of the principle of conservation of energy. Such a violation we shall henceforth refer to by the symbol $\chi$ (from $\chi\eta\rho\alpha\mu o\varsigma$, a *cleft* or *gap*). The point of interest here is that $\chi$ should be empirically detectable. When human beings are asked to consider Argument A, and then decide if they are conscious, then -- assuming all human beings are conscious, and know it -- $\chi$ should be detectable within their brains.

**Single-mind Q may partially conceal $\chi$**

I hope to examine the problems associated with the detection of $\chi$ in future work. However it is first necessary to examine a possibility which may make $\chi$ intrinsically undetectable, at least under certain conditions. This section will require a small diversion into quantum mechanics. In the most common interpretation of quantum mechanics (the Copenhagen interpretation) the physical state of a quantum system is represented by a vector in Hilbert space (Von Neumann 1955). This state evolves deterministically according to the unitary dynamics of quantum mechanics (Barrett 1999). Measurements are represented by applying an appropriate operator (in the form of a Hermitian matrix) to the state vector, which produces a representation of the state vector in terms of some particular measurement basis. The physical states represented by the measurement basis are called eigenstates, since these are the states which result when the state vector is an eigenvector of the corresponding operator. Normally, however, the state vector will be a superposition of basis states, and on measurement this vector is assumed to "project" non-deterministically to an eigenstate of the measurement basis. This is the well-known "collapse" or "reduction" of the state vector.

The problem is that quantum mechanical theory does not provide any clear explanation of what constitutes a measurement. In order to circumvent this difficulty, must supervene on some physical process, whereupon Argument A proceeds as before.

attention has focused recently on so-called "no-collapse" interpretations, in which the physical state never collapses and superpositions persist indefinitely (see Barrett, 1999 for a review). However this now presents us with another problem; how to account for the determinate nature of our experiences, which are always of single "classical" properties and never of superpositions of states. One approach to dealing with this is the single mind Q interpretation (Barrett 1999)[5]. Single mind Q assumes some particular property Q, which evolves in such a way as to ensure that all our experiences are determinate. But in this approach, Q is regarded as a purely mental property -- single mind Q, in other words, postulates a robust mind-body dualism. Q also functions to orchestrate or co-ordinate the experiences of different minds; without this, different minds would experience completely different and potentially unconnected realities.

The consequences of this for detecting $\chi$ are as follows. The process of neuroscientific inquiry can be regarded as the partitioning of a set U, which contains every possible neural topography. Each element of U -- that is, each neural topography -- is a fully specified set of neurological properties (or as fully specified as quantum indeterminacy will allow). The term "neurological properties" is here intended to refer to *all* brain properties, and not necessarily just positional ones. The partitioning of U will yield a subset which I shall call N. As neuroscientific inquiry advances, the set N would be expected to get smaller and smaller.[6]

However in a no-collapse theory, the physical state of the brain underlying U is assumed always to be a quantum state. It is important to be clear about what is going on here. The elements of U are not themselves quantum states. In fact in single mind Q, they are not really physical states at all. They are best understood as classical *appearances*; that is, they are descriptions of how neural topographies appear to the neuroscientists who are observing them. They are purely mental properties. (The determinate nature of these experiences is guaranteed by the determinate property Q, which is a property of the combined system of observer plus brain being studied.)

There are two ways in which U can be partitioned. First, as the physical system evolves, correlations will develop both between neurological properties and other neurological properties, and between neurological properties and properties in the environment. As this evolution occurs some elements of U will become inconsistent with the physical state. One can say that these elements are *partitioned out* of U, and not included in N. The second way U can be partitioned is via the process of quantum measurement; that is, the selection of an eigenstate for some observable. Since in single mind Q this sort of partitioning is always a mental process, the physical state remains unchanged after each partition. However the net effect of both types of process is to produce a subset N which is smaller than it was before.

---

[5] Single mind Q is in fact just one example of a type of theory called a "Q theory". In other versions of Q theory, Q is regarded simply as a physical parameter. In these versions of Q theory mind-body dualism is obviously not required.

[6] The technically-minded reader will have noticed that this is somewhat oversimplified. Although the classical requirement that neuroscientific inquiry is possible ensures that the subset N will reduce in size over time, quantum indeterminacy means that it will not do so smoothly; individual elements of U will "jump" in and out of N as N is refined. The reason for this apparent anomaly is that, in order to keep the representation simple, I have deliberately ignored the difference between static topographies -- those defined at some precise instant of time -- and dynamic topographies, which evolve over time. Neuroscientists who aim to understand the brain are typically interested in dynamic topographies. If one assumes that quantum mechanics plays no functional role in neural processing, then the dynamic topographies can be considered as evolving in essentially classical ways. In this case the quantum indeterminacy in the static topographies can be considered as noise and disregarded. From a neuroscientist's perspective, the physical state can therefore be regarded as a set of classical topographies which is subsequently partitioned by measurement.

There is here a potential loophole by which the effect $\chi$ might be partially concealed. Consider how the brain is normally thought to function; it is a physical system which instantiates what might be called *intelligent processes*. These are processes which enable the brain to respond to a wide range of environmental stimuli without requiring a separate programmed behavior for each stimulus. (Assumptions 1 and 2 may be regarded as an operational definition for such intelligent processes in human beings.) The understanding of these processes is the business of the so-called "special" sciences, such as psychology and cognitive neuroscience. Intelligent processes are assumed to supervene on the physical processes which instantiate them.

Now consider the following thought experiment. Imagine an enormously powerful oracle, which is able to give accurate and meaningful answers to every question asked of it. Such an oracle would appear omniscient to all those by whom it was questioned. But consider that the actual number of questions such an oracle is likely to be asked in a finite period of time is probably a very small fraction of the number of questions which *could* be asked. If it were possible for the oracle to know in advance which questions would be asked, then the oracle could perhaps contrive to know the answers to just those questions and not trouble itself about those questions which no-one would ask. The oracle would still appear omniscient to all those who questioned it; but in practice it would be no such thing.

An analogous situation potentially exists in the relationship between intelligent processes and the physical processes which instantiate them. Of course no-one believes that intelligent processes are all-powerful, but they are very likely far more powerful then is needed to deal with the whole range of situations which arise within a given human lifetime. That is, intelligent processes are capable of dealing with many situations that never in fact arise. This is assumed to be necessary because no-one can predict what situations will actually arise within a human lifetime, even though most of them will never occur. But what if the actual state of the brain were indeterminate at the moment each novel environmental situation arose? In that case the conscious experience of each new environmental stimulus could be regarded as a further partitioning of N. If the actual state of the brain were indeterminate then the resulting partition would contain all neural topographies consistent with the correct response to that stimulus (except those which had previously been partitioned out of N). In most cases this would include all topographies which fully instantiated intelligent processes, but would also include many topographies in which intelligent processes were only partially instantiated (because these topographies would not yet have been partitioned out by measurement).

In the previous section it was shown that any physical system which fully instantiates the human capacity for deductive reasoning will be unable to conclude with certainty that it is conscious (or indeed that it has any other property). But this does not necessarily apply to systems which only *partially* instantiate human deductive reasoning. How does this work in practice? Successive neurological observations and conscious experiences will partition the set U and Q will evolve to ensure the partition is determinate. But quantum measurement will partition U in such a way as to select precisely those topographies which are consistent with those observations and experiences, as long as such states are available -- that is, as long as topographies which are consistent with those observations and experiences remain in N. For example, consider a neural topography which contains a population of cells whose only purpose is to force M to answer "yes" whenever the question "Am I conscious?" is asked. The previous section showed that such a topography could not be consistent with intelligent processes which incorporate a capacity for honest deductive reasoning. But so long as such topographies remain within N, then quantum measurement will select precisely those topographies, and these topographies could be observed through neurological research. Indeed those topographies which *did* fully instantiate intelligent processes would be inconsistent with the conscious experience of self-certainty, and would therefore be selected out by the partition and hence *not included* in N. So the price one pays for consistency between conscious states and physical states is a lack of consistency between the selected topographies and the intelligent processes which supposedly supervene on them. One can see that in such a case the effect $\chi$ would not occur.

Of course in practice it is not just the particular sample of environmental situations which occur within a given human lifetime which one has to consider, but the sample of such situations which occur throughout the whole of human evolutionary history. As the range of actual environmental situations encountered by human beings throughout history becomes larger and larger, the permissible deviation of the topographies in N from perfect consistency with intelligent processes becomes smaller and smaller -- just as, in the case of the oracle, as the number of questions actually asked of the oracle gets ever larger, the oracle will have to get ever closer to true omniscience.

There are two potential difficulties with using single mind Q to "conceal" the effect $\chi$. Firstly, mental operations such as deciding that one is conscious are not really like measurements of quantum observables. In the measurement of a quantum observable an eigenstate of that observable is selected randomly, in accordance with the quantum amplitudes associated with the various eigenstates. But in the specific example of deciding that one is conscious, only those neural states which are consistent with the outcome of that process are possible. Correlation of the observer's physical state with the observer's own mental state removes any possibility of quantum indeterminacy in this particular case.

Since clearly we must be correlated with our own brains this presents no problem for us. But consider an extraterrestrial visitor who is *not* correlated with our brain states or our mental states. Such a visitor would find it extremely peculiar that the usual rules of quantum indeterminacy were being flouted. One can see why by considering the example above of a population of cells whose sole purpose is to ensure that we always answer "yes" whenever the question "Am I conscious?" is asked. Such a neural topography, and the evolutionary history leading up to it, would be extremely unusual. An extraterrestrial visitor uncorrelated with our mental and brain states would expect to find many examples elsewhere in the universe of conscious beings whose brains did not exhibit such a topography. We would thus be unusual in being perhaps the only conscious beings in the universe who could be certain of being conscious, a circumstance which appears unreasonable.

One way round this problem would be to require that all intelligences in the universe, including all extraterrestrial intelligences, were in fact correlated with our own mental and brain states in some fundamental way. The source of such a correlation would presumably have to be found in the very early history of the universe. Another way would be to impose a requirement that the "minds" in single mind Q entail certain properties, and to require that the neural topographies they select be fully consistent with intelligent processes. In this second case the rules of quantum indeterminacy could be preserved, and $\chi$ would not be concealed and should be detectable.

The second problem is that single-mind Q in any case would not *completely* eliminate the possibility of $\chi$. Consider a comprehensive program of neuroscientific research, as represented by a long sequence of measurements, completed before any attempt was made to detect $\chi$. The result would be a subset V, which would be the intersection of all those subsets of U selected by their respective measurements. If the research program were intensive enough then V might be a very small subset indeed. In such a case, one could not be sure that V would still contain sufficient neural topographies, that at least one would remain which was consistent with the mental property of knowing that one is conscious. All neural topographies consistent with that outcome might have been partitioned out by the previous sequence of measurements. In such a case one would expect $\chi$ to be detectable subsequently.

Note that the subset V can be defined as follows:

$$V = U \setminus (V_n \cup V_e \cup W)$$

where $V_n$ is the subset of U inconsistent with neuroscientific observations; $V_e$ is the subset of U inconsistent with observed environmental properties; and W is the subset of U inconsistent with the existence of the non-physical "minds" required by single mind Q. The considerations in this section can be summarized by saying that, if the correct quantum statistics are to be maintained, then either all "minds" in the universe are correlated, or "minds" which are certain of their existence are found only on earth, or it is the case that the subset W is not empty.

**Single mind Q may explain a specific operational definition of free will**

There is a sense in which the single mind Q approach to quantum mechanics may explain a certain notion of free will. To see how this is so, we must now refer to some celebrated experiments by the physiologist Benjamin Libet. The first experiments of interest here refer to a phenomenon generally known as the *readiness potential* (Libet, Gleason, Wright and Pearl 1983; Libet 1985). When human subjects are asked to time as accurately as possible when they experience the impulse to perform a random movement, an EEG trace is observable up to 0.3 seconds before the subject's first conscious awareness of the impulse (this number is an average computed from aggregate data). This is known as the readiness potential. It might be argued that, since the EEG trace precedes the conscious impulse and in effect predicts it, the apparently random conscious impulse is not, in fact, random at all but determined by the neurophysiological state of the subject's brain. So, to the extent that one regards random impulses as a matter of free will, Libet's results can be taken as an argument against free will.

Libet's interpretation of these findings is controversial, particularly with respect to the readiness potential; and it is not my intention here to attempt to resolve this controversy. I wish to make the much narrower point, that even if the readiness potential can be regarded as a predictor of the subject's decision in a classical system, it cannot necessarily be regarded as such in a quantum system. The reason is that the neurological properties underlying the readiness potential may not actually have determinate values until the subject becomes consciously aware of their decision. In connection with this, an earlier experiment (Libet, Alberts, Wright and Feinstein 1972) is of interest here. Using a technique known as *backward masking* which, for reasons of space, will not be described here, Libet found evidence that perceptual stimuli can take up to 0.5 seconds (with a minimum of 0.4 seconds) before they register as conscious impressions -- it takes that long for the subject's brain to process them. This delay is called *perceptual latency.*

Single mind Q illustrates how the second effect may counteract the first. Consider an EEG machine which is in a superposition of two states; a state $EEG_{ON}$, in which the readiness potential is detected, and a state $EEG_{OFF}$ in which no readiness potential is detected. These states are correlated with brain states $BRAIN_{ON}$ and $BRAIN_{OFF}$, in which the readiness potential occurs and does not occur respectively. From the perceptual latency effect described above, it will take roughly 0.5 seconds for the states $EEG_{ON}$ and $EEG_{OFF}$ to form a conscious impression in the mind of the observer reading the EEG machine-- at which point, according to single mind Q, the superposition will be resolved to a single determinate state (albeit only in the minds of the conscious observers). But by that time, the subject's conscious awareness will already have selected a determinate value for the readiness potential, since the readiness potential is shorter than the perceptual latency.

In other words, it is impossible for any observer to perceive consciously if a readiness potential has in fact occurred, *before* the experimental subject experiences the conscious impression of a random impulse. Since in single mind Q determinate properties are mental properties, this means there simply *is no* determinate state for the readiness potential *or* the EEG trace before the subject becomes aware of their conscious decision. The readiness potential therefore cannot, *even in principle*, be used to predict the subject's decision before it happens. This will always be the case if the perceptual latency is longer than the readiness potential. And so, according to single mind Q, it will be the subject who determines the state $BRAIN_{ON}$ or $BRAIN_{OFF}$, and hence the state $EEG_{ON}$ or $EEG_{OFF}$, by random selection. This state of affairs is empirically indistinguishable from the operational definition of free will posited by Libet, but removes any possibility that the readiness potential can be said to have a determinate value before the subject's conscious decision. Of course, this only applies to the rather limited sense of free will described by Libet. It is also subject to empirical review should subsequent research challenge the relative values of the readiness potential and perceptual latency.

What sort of neural mechanism might be implied by the effect described here? A neural network which exploits single mind Q might have the following properties: P is a population of cells, and $I_1$ and $I_2$ are, respectively, excitatory and inhibitory inputs to P. X is a population of cells I shall call the *state determiner* -- Population X determines the output of the network. E and Y are populations which are connected to X by reciprocal excitatory and reciprocal inhibitory connections respectively. X is connected to P by a delay line, which allows small changes in P to manifest before they are amplified by the connections from X to E and Y. $K(P)$ is the mean activity level[7] of P the value of which is equal to $K_{idle}$ when $I_1 = I_2$. The network is set so that when the activity level of X is $K_{idle}$, both E and Y are inactive. An increase in the activity level $K(X)$ of X will drive $K(X)$ to a level $K_{max}$, and a decrease will drive $K(X)$ to $K_{min}$, which are respectively the maximum and minimum values of $K(X)$.

We now introduce a quantum noise term[8] $\varepsilon$ to P. (It is important to note that merely adding classical noise to the network will not work, since the effect being exploited here relies on the quantum superposition being maintained until a conscious decision is made.) We assume $\varepsilon$ to be approximately Gaussian in distribution, with a mean of zero. Therefore when $I_1 = I_2$, the activity level of P will be:

$$K(P) = K_{idle} + \varepsilon$$

The effect of this is to introduce a small variation in $K(P)$ which will quickly be amplified by the network so that the state determiner X will evolve to either $K_{max}$ or $K_{min}$. In quantum mechanical terms, the state vector of the network can be represented as a superposition of two states: a state MAX in which $K(P) = K_{max}$, and a state MIN in which $K(P) = K_{min}$. According to single mind Q, a single state, either MAX or MIN, will then be selected randomly once a conscious observation is made. (Different probabilities for $K_{max}$ and $K_{min}$ can be arranged by varying $I_1$ and $I_2$ so that $K(P)$ is initially either slightly greater or slightly less than $K_{idle}$).

**Consciousness as a fundamental entity in explanations of nature**

Finally I want to make a brief remark about how theories of consciousness, and its

---

[7] Each cell in P, X, E and Y fires a number of action potentials within a certain time $\Delta t$. This number is assumed to follow a Poisson distribution with mean $\mu_K$. Excitatory or inhibitory inputs are assumed to increase or decrease the value of $\mu_K$.

[8] The most likely source of such noise is thought to be in the random variation of neurotransmitter release at neural synapses (Destexhe 2012). If these small random variations are considered equally likely to increase or decrease the likelihood that a cell will fire an action potential, then the cumulative effect of many such variations can be regarded as Gaussian distributed with a mean of zero, if the number of effects is sufficiently large. It is unfortunately impossible to quantify these effects in any simple way since they depend critically on the internal connectivity of the network, and in particular on the extent of feedback connections within the populations of cells.

interaction with the physical world, should include consciousness itself as an entity. Since consciousness cannot be fully decomposed into physical components, how can it be defined as a theoretical entity, and what properties should be attributed to it? The obvious starting point is to define consciousness in terms of precisely that property which turns out to be
inconsistent with physical decomposition -- that is to say, self-certainty. This property can be defined in terms of the mapping E → X which was set out in the first section of this article. If we refer to this mapping as the function $p_0$, then self-certainty can be defined as the capacity of consciousness to perform the function $p_0$ with provable reliability and accuracy. This can be defined symbolically in terms of an infinite sequence of functions:

    $p_1, p_2, p_3, \ldots$

where every $p_n$ can be defined in the following terms:

    $X_n = p_n(p_{n-1})$

such that:

    $X_n$ = YES if $p_{n-1}$ is performed accurately and reliably;

    $X_n$ = NO otherwise.

Clearly, each function $p_n$ in the sequence examines whether the previous function $p_{n-1}$ has been correctly performed. These functions obviously correspond to the computations (or physical processes) described as part of the infinite regress in Argument A. However, unlike those processes, these functions are merely abstract representations of the *properties* of consciousness, and are not concrete entities in the physical world. In fact the representation of self-certainty in terms of a sequence of functions provides another way of proving the impossibility of self-certainty in a purely physical system, since it is easy to show that no physical system can perform all of these functions. To see why, one need only assume some physical process $P_n$ which performs each function $p_n$. If one assumes the functions are performed sequentially, then one notes that each $P_n$ requires some time to execute, say $\delta t$. The infinite sequence of functions therefore requires a total time of $\delta t$ multiplied by infinity. Alternatively if one assumes the various functions are performed in parallel, then each $P_n$ requires some region of space, say $\delta V$, to execute. The total volume of space required to perform all the functions simultaneously is therefore $\delta V$ multiplied by infinity. A physical system to perform the infinite sequence of functions would therefore need either to be infinitely large or to take an infinite amount of time, and neither contingency is physically reasonable.

The infinite sequence of functions can be summarized as a single function $p_\omega$, identified by the subscript ω or *omega*:

    $X_\omega = p_\omega(E)$

where:

$X_\omega$ = YES if it is provably the case both that E = 1 and $p_\omega$ is reliably performed;

$X_\omega$ = NO (or more accurately, is undefined) otherwise.

This is purely a notational convenience. One can regard a defining characteristic of consciousness as the ability to perform the function $p_\omega$, and a defining physical property of consciousness as the $\chi$ effect (or violation of energy conservation) which is associated with it. Once defined, such a fundamental entity can be included in theoretical models, or simulations, of neurological or cognitive processes. This illustrates that it is not true, as is sometimes claimed, that allowing a non-physical basis for consciousness renders it immune to analysis or understanding.